# Propagation analysis and prediction of the COVID-19

Lixiang Li, Zihang Yang, Zhongkai Dang, Cui Meng, Jingze Huang, Hao Tian Meng,

Deyu Wang, Guanhua Chen, Jiaxuan Zhang, Haipeng Peng.

Corresponding author: penghaipeng@bupt.edu.cn

Information Security Center, State Key Laboratory of Networking and Switching Technology, Beijing University of Posts and Telecommunications, Beijing 100876, China

**Abstract:** Based on the official data modeling, this paper studies the transmission process of the Corona Virus Disease 2019 (COVID-19). The error between the model and the official data curve is within 3%. At the same time, it realized forward prediction and backward inference of the epidemic situation, and the relevant analysis help relevant countries to make decisions.

## Main results:

- ✓ **Analysis of the epidemic situation in Hubei:**

1). When will more than 10,000 confirmed coronavirus cases be completely cured in Hubei?

Answer: The model predicts that almost all will be cured on April 1. (Fig. 1)

2). When was the earliest case in Hubei?

Answer: We find that there was already an infection on November 24. (Fig. 2)

3). How many people are infected per person (basic reproduction number)?

Answer: It is 3.8 when not under control, 0.5 after closing Wuhan city and 0.1 after closing of Wuhan community (Fig. 3)

4). What is the average incubation period of the virus?

Answer: 6 days. (Fig. 4)

5). If it is controlled 5 days in advance, how much is the infection? What if control is lagging 5 days?

Answer: If 5 days in advance, the number of infected people will be 28000, 42% of the current number of confirmed cases. If the control is delayed for 5 days, it will reach 156000 people, 2.26 times of the current level. (Currently diagnosed 68000 in Hubei) (Fig. 5)

6). What distribution does the daily infection curve satisfy, which day reaches the peak, is February 12?

Answer: Normal distribution, the actual peak appeared on February 8. In the official data, because the clinical diagnosis was not added before the February 12, the data surged on the February 12 (jump to 14,840 people, cumulative results), indicating that the previously published data did not reflect the actual infection

situation. (Fig. 6)

7). The average number of days from diagnosis to cure?

Answer: In the stage of Hubei epidemic, it takes an average of 21 days for patients from diagnosis to cure. (Fig. 1)

✓ **Epidemic situation in non-Hubei areas:**

8). What is the law of transmission in non-Hubei areas?

Answer: The epidemic transmission curve of non-Hubei area is similar to that of Hubei area, which was controlled 10 days in advance (on January 13). (Fig.7)

✓ **International epidemic situation except China**

9). How to predict the epidemic situation in South Korea?

Answer: At present, the epidemic situation in South Korea is basically under control. According to the model, South Korea will be basically under control by the end of March. It was found on January 7 that there was infection in South Korea (confirmed by official broadcast January 20. In addition, we found that the basic reproduction number before the control was 4.2 in Korea and 0.1 after the control. (Fig. 8)

10). How to predict the development of Italian epidemic?

Answer: At present, 15000 people in Italy are infected. If not controlled, it will increase dramatically. It will reach 200000 by the end of March. In fact, Italy began to control on March 8. According to China's basic reproduction number (0.1) and South Korea's basic reproduction number (0.1), it will reach 84000 by the end of March. At present, the basic reproduction number in Italy is 4.2. According to the model inversion, the infection was found on January 13 in Italy (2 cases were confirmed by the official broadcast January 31). (Fig. 9)

11). How to predict the development of the epidemic in Iran?

Answer: At present, 11,000 people are infected in Iran. Our model predicts that it will reach 20,000 people at the end of March, through backward inference of the epidemic situation, we found that Iran had an infection on January 13 (officially reporting that two cases died on February 20, and there was no official data before). The model found that the basic reproduction number in Iran before the control was 4.0 and after the control was 0.2. (The Iranian government took many measures in early March.) (Fig. 10)

## 1 Specific analysis of the epidemic situation model

The epidemic data of Hubei Province is large [1], more in line with the statistical law, and the statistical model is more able to reflect the process of virus transmission, so we build a statistical model for the epidemic situation of Hubei Province (see the method section later for details). Based on the data provided by National Health Commission of China(NHCC), we have carried out accurate simulation. By comparing the simulation results with the real data, we analyze the propagation process and its influencing factors.

## 1.1 Simulation comparison of epidemic model in Hubei Province

According to the comparison between the official epidemic data and simulation data of Hubei Province (see Figure 1 for details), we can see that the model simulation curve of the number of confirmed infections, the number of cured people and the number of dead people matches the official data curve very well. According to the curve of simulated number of infected persons and the curve of official number of infected persons, we can see that there is a certain gap between the number of infected persons before February 12, and the others are basically the same. (after the official announced the method of clinical diagnosis of pneumonia on February 12, the data on that day surged to 14000, indicating that the official data had some omissions before February 12, and our model give a more accurate response to the number of actual cases).

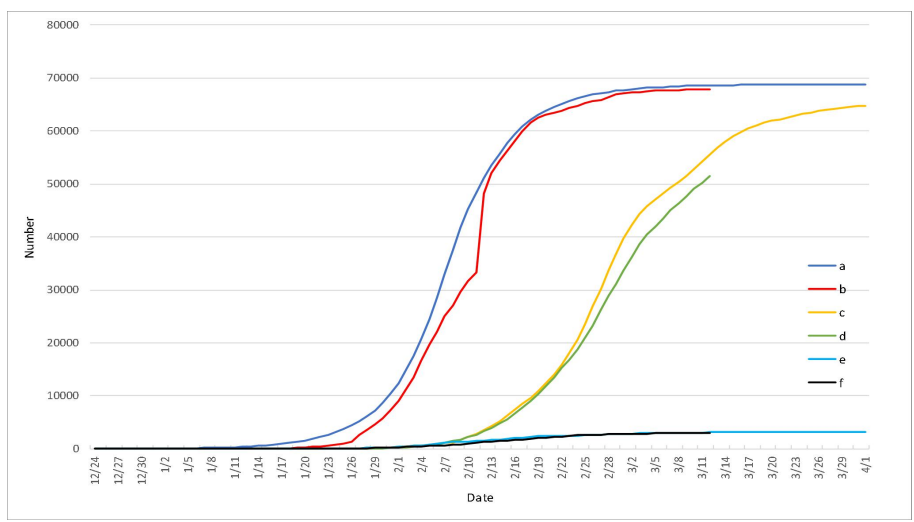

**Fig.1.** The comparison between the official epidemic data and simulation data of Hubei Province (The curve a - Number of simulated infections. The curve b - Number of officially confirmed infections. The curve c - Number of simulated cures. The curve d - Number of officially cures. The curve e - Number of simulated deaths. The curve f- Number of officially deaths.) There are three stages in the simulation. The first stage is before January 23, when Wuhan is not closed; the second stage is from January 23 to February 10; the third stage is after February 10, when Wuhan starts to close the community.

## 1.2 Simulation of the number of infected people

According to the curve of simulated infection number, we infer that by the end of

March (yellow line + light blue line = dark blue line：Number of simulated cures+ Number of simulated deaths= Number of simulated infections), all cases will be basically treated, that is to say, more than 10000 cases in Hubei will be basically cured and cleared. Note: the number of cases of infection on December 24 is not zero. Because the data in Figure 1 is too large, small data cannot be displayed. For this reason, according to the same distribution curve, we also give the simulation data results before December 24, and the simulation found that the initial infected people had existed as early as November 24 (see Figure 2 for details). The first confirmed infection was reported on December 8 [2].

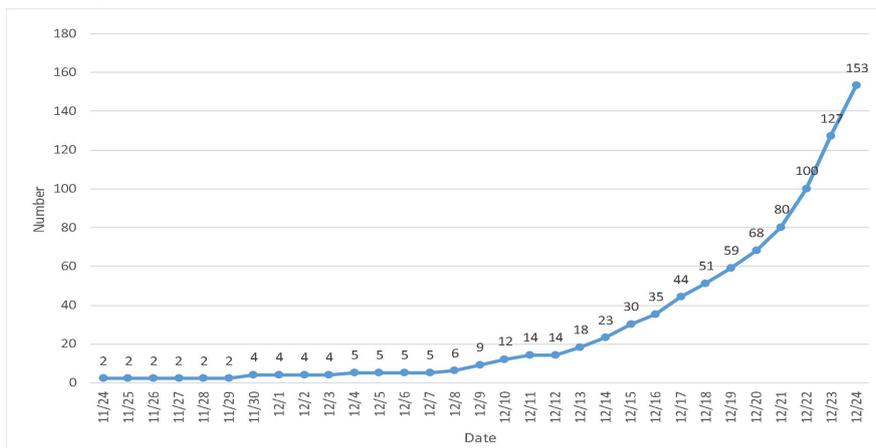

**Fig.2.** Backward graph of initial epidemic curve.

1.3 Simulation of the number of the average infected people

The basic regeneration number has a great influence on the spread of the outbreak. As the number of basic regeneration increases, the total number of people infected will also increase; and the greater the number of basic regeneration, the faster the disease will spread. As in the previous simulation, there are three phases. The first phase is an uncontrolled phase. Through a large number of simulations, we found that the basic regeneration number is 3.8, and the second and third phases are controlled phases. The basic regeneration numbers are respectively It is 0.5 and 0.1 (see the yellow line in Figure 3). The simulated curve (yellow line) can be closer to the official diagnosis curve (red line) in Hubei. The first stage is the free propagation stage, and we also give other simulation results of the basic regeneration number in the range of 3-4.6 (the second and third stages remain unchanged). The effect of transmission is significant, with the most accurate final infections being obtained only at 3.8. As for the basic regeneration number in the stage of free propagation, some scholars have proposed that it should be between (2.8, 3.9) and 2.2. Obviously, these expressions are either inaccurate or too wide.

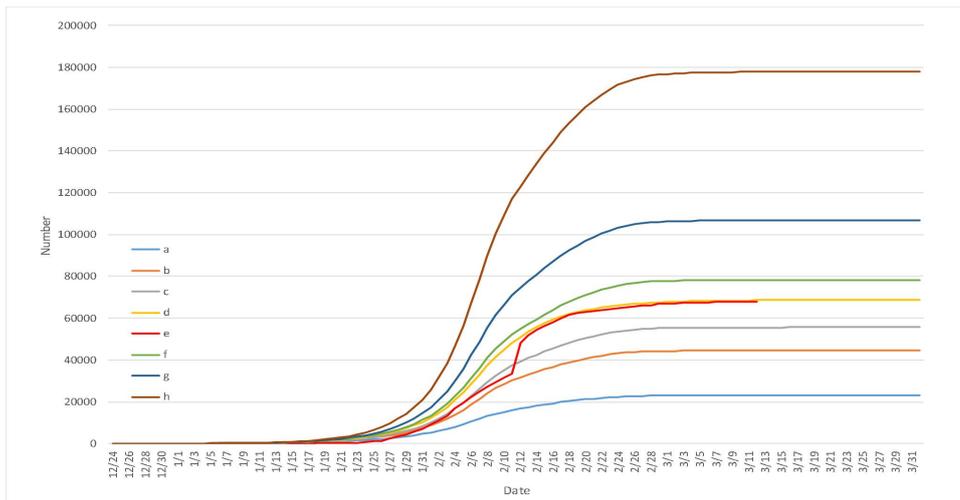

**Fig.3.** Comparison of the basic reproduction number. (The basic reproduction number a = 3, b = 3.5, c = 3.7, d = 3.8, f = 3.9, g=4.1, h=4.6, respectively. The curve e - Number of officially confirmed infections.)

### 1.4 Simulation of the average latency

Figure 4 shows the comparison chart of the average incubation period. We simulated the days in the range of 3-9 respectively, and found that when the average incubation period was set to 6 days, it could better fit the official curve, while the other simulated values are quite different from the official curves. It is also found that the shorter the incubation period, the faster the virus spreads and the greater the total number of infections. David mentioned that the incubation period of the virus is 3-6 days, which is not accurate.

According to the previous analysis, the disease develops on average after 6 days of incubation, and it takes an average of 5 days from the morbidity to the diagnosis (data from CHCC)[4], that is, 5 + 6 = 11 days from the initial infection and incubation to the diagnosis and isolation after the morbidity. Obviously, not all 11 days are contagious. Simulations show that when the average time for contagion is set to 8 days, it fits the official curve best.

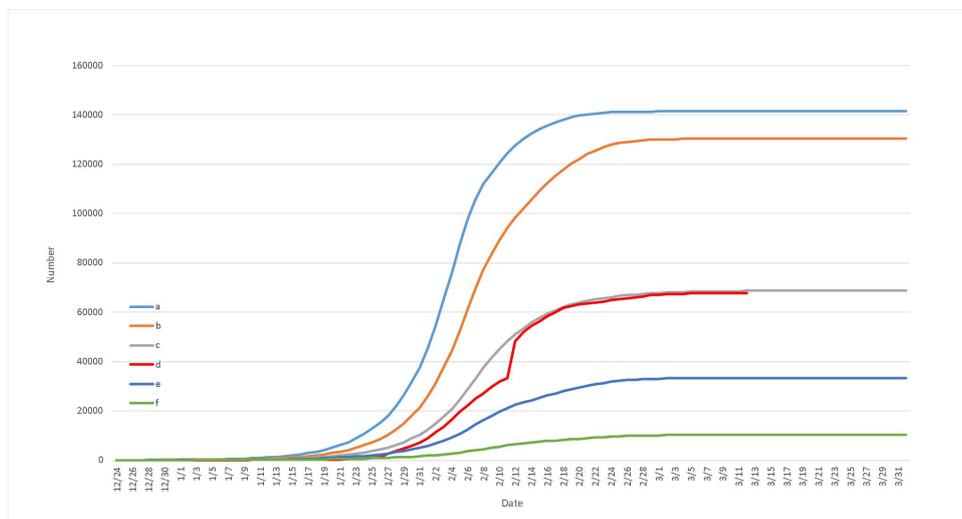

**Fig.4.** Comparison of the average latency. (The average latency a = 3, b = 5, c = 6, e = 7, f = 9, respectively. The curve d - Number of officially confirmed infections.)

1.5 The simulation test of the epidemic control

We know that epidemic control at different times (such as closing city) will have an important impact on the spread of the epidemic, obviously the earlier the better. Combined with the actual situation of Wuhan's closure from January 23, we simulated the comparison of the spread of the epidemic situation on January 18 and January 28, as shown in Figure 5. The red curve and the orange curve in the figure respectively represent the official curve and the simulation curve (when the "cities were closed" measures were taken on January 23). Obviously, the two curves match well, and the simulated final number of infected people is 69000, which is also close to the actual situation. If the control is started five days in advance, that is, on January 18, the blue line shows that the number of infected people is about 28000, which is 0.42 times of the number of confirmed cases. If the measures are delayed for 5 days, it can be seen from the yellow line that the number of people will be as high as 156000, about 2.26 times of the current number of patients. Zhong Nanshan's team once predicted that if the closure measures were delayed for five days, the number of patients would reach three times of the current number, i.e. 210000 [5]. Our results are more optimistic. It can be seen from this figure that it is necessary to take timely measures to control the spread of the epidemic, which can also be used as a reference for some overseas countries.

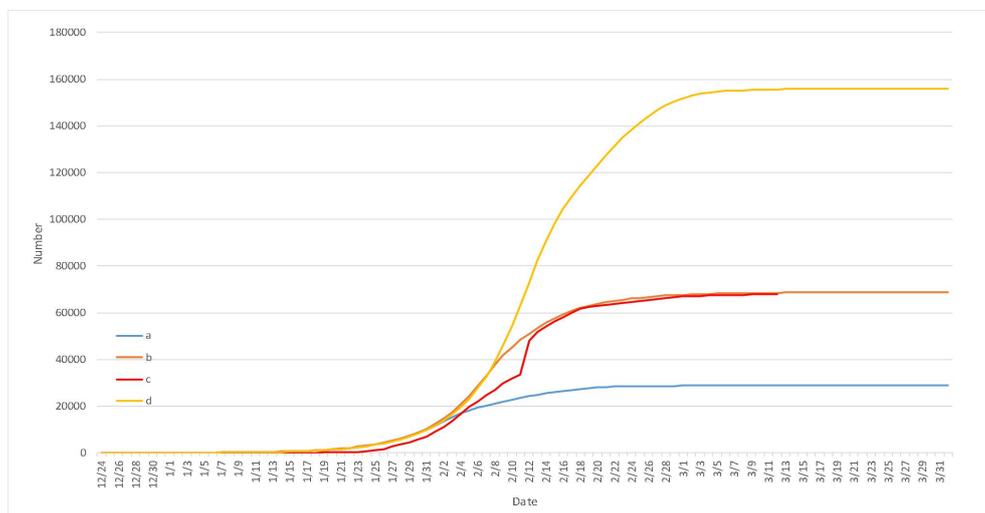

**Fig.5.** Comparison of epidemic control at different times (The curve a - 5 days in advance. The curve b - Control by actual days. The curve c - Number of officially confirmed infections. The curve d - 5 days delayed.)

1.6 The simulation of daily new infections

We also studied the curve of the daily increase in the number of people (Figure

6). Because the official did not include the imaging features of pneumonia in the clinical diagnosis into the diagnosis conditions for statistics before February 12, 2020, the omission and inaccuracy of the previous data caused the official data to jump on the February 12 (the number of newly diagnosed patients in Hubei was 14840 people).As shown in Figure 6, our simulation data meet the normal distribution, at this time, the daily new infection data peaked on February 8, with the number of 4500. However, the published data, due to not timely included in the clinical diagnosis image test (red line), many patients were not diagnosed, but the curve has become a downward trend, data deviation. For this reason, we smoothed the clinical diagnosis data on February 12, 13 and 14, and used the Gauss function of latent period. These three days were chosen because of the sudden increase of the number of clinical diagnosis in these three days. It can be seen from the figure that the smooth number of official daily infected persons (black line) and the simulation curve (blue line) fit very well.

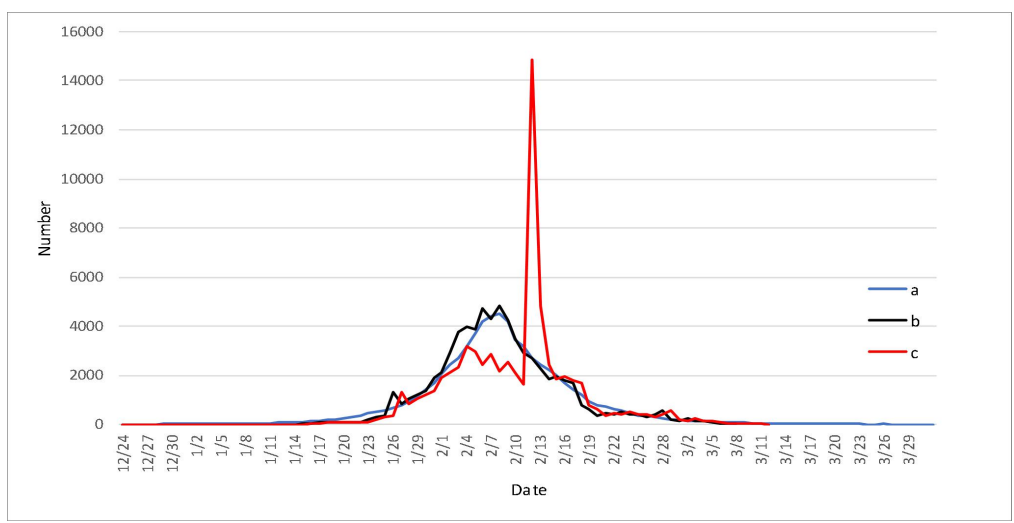

**Fig.6.** Change curve of daily new infections. (The curve a - Simulated curve. The curve b - Smoothed curve. The curve c - Number of officially confirmed infections.)

## 2 Data analysis of non-Hubei regions

Due to the large liquidity before the Spring Festival in China, Hubei Province has produced case output. We have also conducted research on non-Hubei area, which is also divided into three stages: the first stage is before January 26 (30 provinces and cities in the country enter the first level of combat readiness on January 26), the second stage is from January 26 to February 5 (the whole country starts to close the community in early February, here we take No. 5 for simulation), the third stage is after February 5. A large number of simulations found that the regeneration numbers in the three stages were 3.8, 0.5, and 0.1, respectively. At this time, the simulation curve (blue line) and the actual curve (red line) fit well. Another interesting phenomenon is that we found that if Hubei 's transmission was controlled 10 days in advance (Hubei began to control on January 13), its propagation curve (yellow line) is highly similar to the propagation curve in non-Hubei regions.

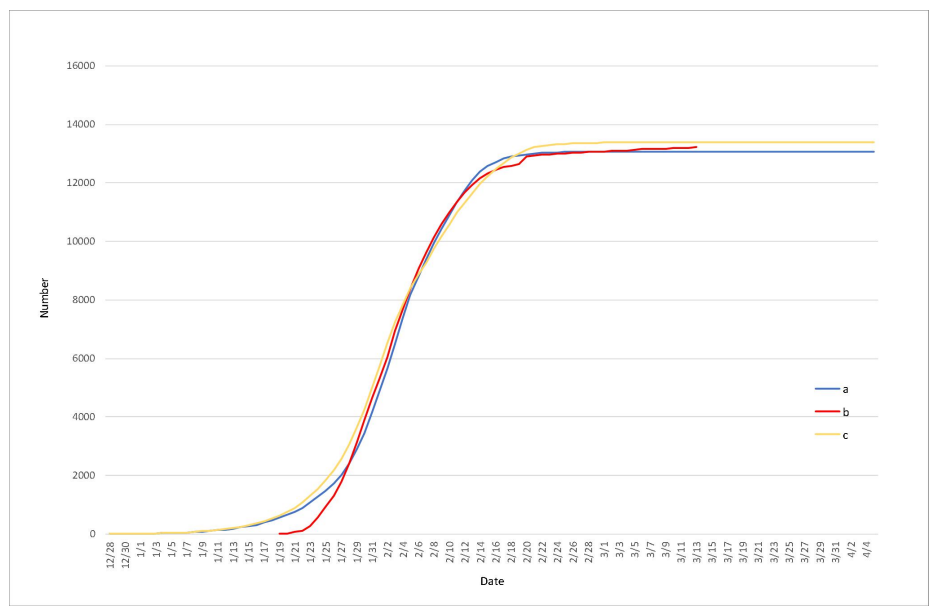

**Fig.7.** Correlation analysis of epidemic data in non-Hubei area (The curve a - Simulated number of non-Hubei infections, the curve b - Actual number of non-Hubei confirmed cases, the curve c - Number of simulated infections in Hubei when control was ten days in advance)

On January 24th, 235 cases were confirmed in non-Hubei province, with an accumulation of 1052 cases in Hubei province, accounting for about 20%. Finally, it was officially announced that non-Hubei 13220 and Hubei 67786 were infected, which was also basically in line with the proportion of 20%. In other words, there is a high degree of similarity between the initial and final transmission ratios in Hubei and non-Hubei areas.

## 3 The Prediction of foreign epidemic

### 3.1 Analysis of Korea data

From March 2, South Korea began to implement large-scale measures to prevent and control the epidemic situation [6]. At present, the epidemic situation in South Korea is basically under control. Our model simulates the actual data well, and it is found through model prediction (Figure 8). That South Korea will be basically controlled by the end of March. We inferred from the curve and found that there was an infection in South Korea on January 7 (the official broadcast confirmed the diagnosis at January 20). We found that before the control, the basic reproduction number of virus transmission in Korea was 4.2, and the basic reproduction number after control was 0.1. (The basic reproduction number reached 4.2. We suspect that some churches in South Korea ignored the spread of the epidemic in the early stage.)

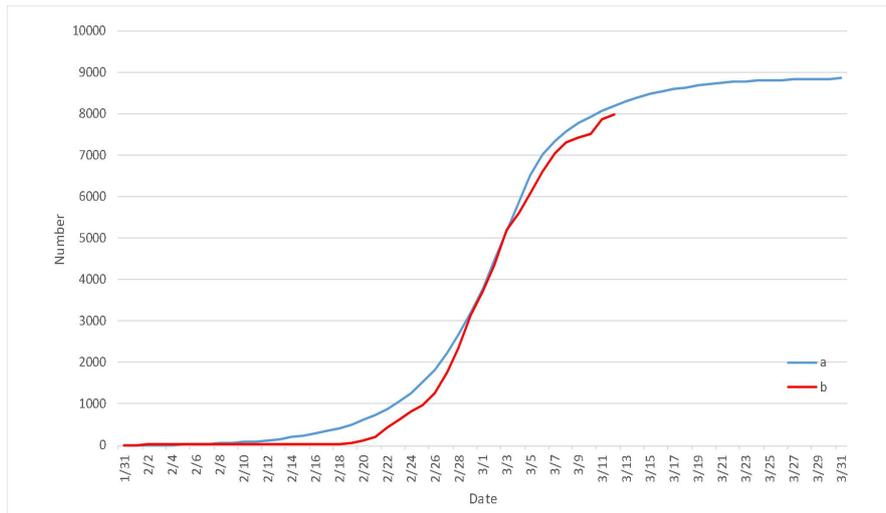

**Fig.8.** Comparison of Korean epidemic simulation data and official data. (The curve a-Number of simulated infections. The curve b-Number of officially confirmed infections.)

### 3.2. Analysis of Italian data

Currently, 15000 people are infected in Italy. Through simulation, we find that (Figure 9), if we do not control (blue line) at all, it will grow explosively, reaching 200000 at the end of March. In fact, Italy began to control population flow on March 8. According to the basic reproduction number (0.1) after China's control and the basic reproduction number (0.1) of South Korea, our model predicts that by the end of March, the number of confirmed cases in Italy will reach 84000 (yellow line). At present, the basic reproduction number in Italy is 4.2. According to the model inversion, there were infected persons in Italy on January 13 (2 cases were confirmed in Italian official broadcast January 29 [7]). (The reproduction number reaches 4.2, which is caused by the lack of attention paid by the public to the epidemic in the early stage.)

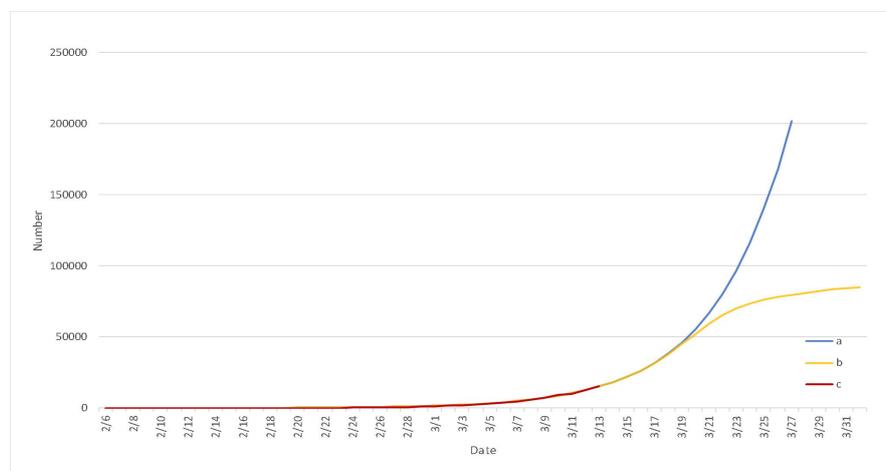

**Fig.9.** Comparison of Italy epidemic simulation data and official data. (The curve a-Number of simulated infections(uncontrol). The curve b-Number of simulated

infections (partial control). The curve c-Number of officially confirmed infections.)

### 3.3 Analysis of Iranian data

Currently, 11000 people are infected in Iran. According to the simulation, as shown in Figure 10, the infection data is expected to reach 20000 by the end of March, and then it will be basically controlled by the beginning of April. It was found that Iran had infection on January 13 (Officially announced 2 cases died on February 20, no previous official data) [8]. The model found that the basic reproduction number of virus transmission in Iran before control was 4.0, and the basic reproduction number after control was 0.2. (the Iranian government took many control measures in early March. The basic reproduction number after control is 0.2, indicating that the control effect is not ideal.)

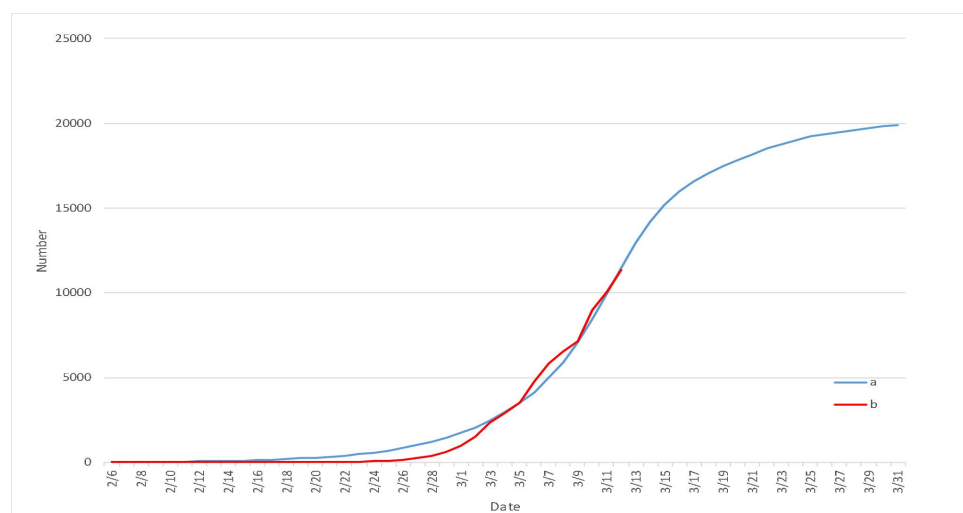

**Fig.10.** Comparison of Iran epidemic simulation data and official data. (The curve a-Number of simulated infections. The curve b-Number of officially confirmed infections.)

## 4 Method and description

### 4.1. Transmission of the virus

Many studies have shown that virological transmission usually satisfies the Gaussian distribution, so this article uses the Gaussian distribution to analyze the transmission of viruses [9-10]. When we use Gaussian distribution to simulate, the main influence is the average value, but the standard deviation has little influence, so we will take the best accord with the objective fact variance to simulate. The basic reproduction number is the number of people infected by a patient during the average illness period when all are susceptible at the morbidity of the disease. The Gaussian distribution function of $D1 \sim N(x1,y1)$ is used to represent the propagation ability of the virus. This formula indicates that a single infected person can infect $x1$ person on average in the first stage. At this stage, x1 has a value range of $(1, +\infty)$, and a standard deviation of $y1 = 1.5$.

With the development of a series of effective measures, the daily increase of the number of infected people also decreased to a certain extent, and the transmission ability of the disease in the population decreased. For the ability to spread disease through this process, we still use the Gaussian distribution. When a single infected person at this stage is able to infect an average of x2 individuals, the formula for the capacity to transmit is, $D2 \sim N(x2, y2)$. At this stage, the value range of x2 is general $(0,1]$, standard deviation of x2 is $y2 = 2 * x2$. Because the medical condition of affected area and the degree of people's activity intention directly decide the basic regeneration number in this stage. The better the health care, the less active the population, the smaller the x2.

In the course of disease transmission, the patient from being infected to the morbidity of disease, we call it incubation period. We assume that it takes x3 days for people to get infected and become aware of the disease, then the incubation period follows $D3 \sim N(x3, y3)$. In this model $x3 = 6.0$, the standard deviation $y3 = 2.0$.

In addition, patients can transmit the virus to other people, so we assume that the average time from infection to transmission is x4 days, which follows $D4 \sim N(x4, y4)$. According to the previous analysis, the disease develops after an average of six days of incubation, and it takes an average of five days from morbidity to diagnosis (according to NHCC), that is, $5 + 6 = 11$ days from the initial incubation of infection to post- morbidity isolation. Clearly, not all 11 days are infectious, and simulations show that setting the average number of infectious days to 8 best fits the official curve. In this model $x4 = 8.0$, the standard deviation $y4 = 1.5$.

## 4.2. Cure the discharge process

The cure time of the patient is the difference between the time of hospital discharge and the time of diagnosis, set to x5. During the epidemic period, the cure time of patients was expressed by Gaussian distribution, that is, $D5 \sim N(x5, y5)$, which means that it takes an average of x5 days for a single individual to be cured and discharged from hospital. During the early and late stages of the epidemic, depending on the medical condition and level of vigilance, there will be small changes to the x5. As the epidemic continued, people's awareness of self-protection became stronger, and hospital treatment became more effective in the late stage of the outbreak, the patient's recovery time became shorter and shorter. In Hubei Province, the average time from diagnosis to cure was 21 days.

## 4.3. Death toll

Relative to the spread of previous diseases, the early mortality rate of this disease is also very high, mainly due to the lack of understanding of the new virus, and secondly, the new crown virus transmission is very strong, leading to the collapse of the medical system in the epidemic area. Due to the continuous improvement of medical conditions and other reasons, the mortality rate of the disease has continued to ease. We assume that the mortality rate is x6, then the mortality distribution follows D6~N(x6,y6). The mortality rate will change in stages depending on medical conditions. According to the CHCC, Hubei Province's current mortality rate is 4.5 percent.

## 5 Conclusion

In this paper, through analyzing the existing data of Hubei epidemic situation [11], the corresponding model is established, and then the simulation is carried out. Here, we studied the main factors affecting the spread of COVID-19, such as the number of basic regenerations, the incubation period and the average number of days of cure. What's more, we predicted the evolution trend of the existing epidemic data, and found that imposing controls would have important impact on the epidemic [12]. In addition, according to the existing data abroad, we also make bold predictions of the epidemic development trends in South Korea, Italy, and Iran, pointing out the possible outbreaks and the corresponding control time, and tracing the earliest transmission dates of countries. Finally, we hope that this article can make some contributions to the world's response to this epidemic and give some references for future research.

## 6 Acknowledgements

We thank Licheng Wang, Ye Tian and Yetao Lu for helpful comments and discussions. In addition, we also thank Junying Liang, Chengwei Tong, Jiaxin Shi, Tao Zhou, Jingyu Sun, Yameng Zhang and Huizhen Li for their contributions in data search. This paper is supported the National Natural Science Foundation of China (Grant Nos. 61771071, 61573067).
**The data sources**: http://sa.sogou.com/new-weball/page/sgs/epidemic
and https://voice.baidu.com/act/newpneumonia/newpneumonia/?from=osari_feed_tab